\documentclass{article}
\usepackage{spconf,amsmath,graphicx}

\usepackage{enumitem}
\usepackage{url}
\usepackage{hyperref}

\title{A comparative study of computational aesthetics}
\name{Dogancan Temel and Ghassan AlRegib}
\address{Center for Signal and Information Processing (CSIP)\\
School of Electrical and Computer Engineering\\
Georgia Institute of Technology, Atlanta, GA, 30332-0250 USA\\
\{cantemel,alregib\}@gatech.edu}
\begin{document}

\onecolumn 

\begin{description}[labelindent=1cm,leftmargin=3cm,style=multiline]

\item[\textbf{Citation}]{D. Temel and G. AlRegib, "A comparative study of computational aesthetics," 2014 IEEE International Conference on Image Processing (ICIP), Paris, 2014, pp. 590-594.} \\

\item[\textbf{DOI}]{\url{https://doi.org/10.1109/ICIP.2014.7025118}} \\

\item[\textbf{Review}]{Date added to IEEE Xplore: 29 January 2015} \\

\item[\textbf{Code/Poster}]{\url{https://ghassanalregib.com/publications/}} \\

\item[\textbf{Bib}] {
@INPROCEEDINGS\{Temel2014\_ICIP,\\ 
author=\{D. Temel and G. AlRegib\},\\ 
booktitle=\{2014 IEEE International Conference on Image Processing (ICIP)\},\\ 
title=\{A comparative study of computational aesthetics\},\\ 
year=\{2014\},\\ 
pages=\{590-594\},\\
doi=\{10.1109/ICIP.2014.7025118\},\\ 
ISSN=\{1522-4880\},\\ 
month=\{Oct\},\}\\
} \\

\item[\textbf{Copyright}]{\textcopyright 2014 IEEE. Personal use of this material is permitted. Permission from IEEE must be obtained for all other uses, in any current or future media, including reprinting/republishing this material for advertising or promotional purposes,
creating new collective works, for resale or redistribution to servers or lists, or reuse of any copyrighted component
of this work in other works. } \\

\item[\textbf{Contact}]{\href{mailto:alregib@gatech.edu}{alregib@gatech.edu}~~~~~~~\url{https://ghassanalregib.com/}\\ \href{mailto:dcantemel@gmail.com}{dcantemel@gmail.com}~~~~~~~\url{http://cantemel.com/}}
\end{description} 

\thispagestyle{empty}
\newpage
\clearpage

\twocolumn

%
\maketitle
\begin{abstract}

Objective metrics model image quality by quantifying image degradations or estimating perceived image quality. However, image quality metrics do not model what makes an image more appealing or beautiful. In order to quantify the aesthetics of an image, we need to take it one step further and model the perception of aesthetics. In this paper, we examine computational aesthetics models that use hand-crafted, generic and hybrid descriptors. We show that generic descriptors can perform as well as state of the art hand-crafted aesthetics models that use global features. However, neither generic nor hand-crafted features is sufficient to model aesthetics when we only use global features without considering spatial composition or distribution. We also follow a visual dictionary approach similar to state of the art methods and show that it performs poorly without the spatial pyramid step.


\end{abstract}
\vspace{-0.1in}
\section{Introduction}
\label{sec:intro}
\vspace{-0.1in}
Everyday, we are exposed to various images and video thanks to Facebook, Flickr, Youtube, Instagram and others. The content in these websites or applications are provided by the users. While uploading the multimedia content, files need to satisfy basic constraints such as format, size and resolution. However, these social media platforms do not assess the quality of multimedia content. Subjective quality evaluation is by far the best way to asses the quality of multimedia. However, it requires extensive amount of time and labor. Therefore, objective quality metrics are used to estimate subjective quality.

Most of the quality assessment methods estimate  the quality by calculating the degradations in the images and videos. Fidelity-based metrics calculate the accuracy of the processed content with respect to the original content whereas structural metrics model the perceived quality of visual data by considering the Human Visual System (HVS). However, fidelity and structure-based metrics are not sufficient to estimate the quality of experience for the end user. We claim in our work that we also need to consider the aesthetics within images and videos. Hence, in our work, we develop image quality measures that incorporate aesthetics as well as other structure-based and statistical models.

The authors in~\cite{Datta2006} proposed a computational approach to study aesthetics in photographic images. They studied aesthetics as a machine learning problem by extracting low-level features based on rules of thumbs in photography, common intuition and rating patterns. In~\cite{Hasler2003}, authors designed a colorfulness index to asses the quality rather than fidelity. Colorfulness index quantifies the colorfulness in natural images to estimate perceived quality by using the color distribution in the CIELab color space. In addition to the colorfulness feature, brightness, contrast, saturation and saliency were also used to asses the beauty rating of videos in \cite{Yildirim2013}. In \cite{Dhar2011}, the authors used composition, content and sky-illumination as high level attributes to predict aesthetics and interestingness.

Instead of focusing on the entire image, the authors in \cite{Luo2008} extracted subject regions using blind motion deblurring \cite{Levin2006} to detect areas that draw the most attention of HVS. The authors in \cite{Tang2013} used hue and scene composition features as global features whereas dark channel, face region and complexity as regional features. Sharpness, colorfulness, luminance, color harmony and blockiness were used in \cite{Moorthy2010} as low-level features to model visual aesthetic appeal. Composition-specific features such as relative foreground position and visual weight ratio were used in \cite{Bhattacharya2010} to asses photo quality and perform semi-automatic enhancement based on visual aesthetics. Instead of hand-crafting features that highly correlate with photographic practices and techniques, the authors in \cite{Marchesotti2011} used generic image descriptors such as GIST \cite{Oliva2001} and BOV (\cite{Csurka2004}, \cite{Sivic2003}) to asses aesthetics of images.

It is not straightforward to describe aesthetics. Therefore, we need to focus on the judgement of subjects. The authors in \cite{Luo2008} generated a video database by selecting videos from YouTube. The database contains 4000 high quality professional movie clips and 4000 low quality amateurish clips. The quality of the videos was assessed per frame basis and the average was computed to asses the quality of videos. The authors also used MSN Live Search to search for images and volunteers ranked the images on a scale between 1 and 5. A photography database with peer-rated aesthetics scores ranging from 1 to 7 was provided in \cite{Photonet}. Similarly, \cite{DPChallange} provided a large photo database with peer ratings based on quality ranging from 1 to 10. Authors in \cite{Datta08} randomly collected large samples from  \cite{Alipr}, \cite{DPChallange} , \cite{Photonet} and \cite{Terragalleria} that were annotated with aesthetics, quality, liking and emotion scores. Around 1 million images crawled by Flickr with textual tags, aesthetics annotations, and EXIF meta-data were provided in \cite{CLEF2010}. A large set of standardized, emotionally-evocative color photographs were provided with a wide range of semantic categories in \cite{IAPS}. Authors in \cite{Murray2012} generated a large scale database with score distributions, semantic and style labels and rich annotations including aesthetics.

In this work, we focus on the binary classification of the images based on aesthetics quality. We examine the descriptors used in the aesthetics image quality  literature  as well as  other commonly used image descriptors.  In section \ref{sec:literature}, we introduce state of the art computational aesthetics descriptors. Geometric descriptors are discussed in section \ref{sec:genericgeometric} and color descriptors are explained in \ref{sec:genericcolor}. We examine hybrid descriptors in section \ref{sec:generichybrid} and visual dictionary approach in section \ref{sec:genericvisual}. We compare the descriptors in section \ref{sec:comparison} and conclude our discussion in section \ref{sec:conclusion}.


\vspace{-0.1in}

\section{Aesthetics Descriptors }
\label{sec:mainbody}
\vspace{-0.1in}
  We can measure the success of the computational models  by how good they can estimate the average subjective scores. From the image sets referred in section \ref{sec:intro}, we use the \texttt{CUHK} database collected by the authors in \cite{Ke2006}. The images in the database are obtained from the photo contest website DPChallenge~\cite{DPChallange} along with the user ratings. From the obtained 60,000 images, the top 10\% images are selected as \texttt{good} and the bottom 10\% images are selected as \texttt{bad} images. Half of the image set is randomly selected to be used for training with labels and the other half is used for the classification tests. In the following sections, we briefly introduce the descriptors and examine their classification performances.

\subsection{State of the art descriptors }
\label{sec:literature}
Ke at al. \cite{Ke2006} designed aesthetics features using spatial distribution of edges, color distribution, hue count, blur, contrast and brightness. Datta et al. \cite{Datta2006}  used exposure of light, colorfulness, saturation, hue, rule of thirds, familiarity measure, wavelet based texture feature, size and aspect ratio, region composition, depth of field and shape convexity to asses image aesthetics. Tong et al. \cite{Tong2004} implemented a black-box approach by generating a set of low-level features and fusing these features using learning algorithms. In addition to extracting global features, Luo and Tang \cite{Luo2008} extracted subject regions to obtain local features of foreground and background. They calculated clarity contrast, lighting, simplicity, composition geometry and color harmony to model the aesthetics. Marchesotti et al. \cite{Marchesotti2011} used generic features instead of hand-crafted features to perform aesthetics-based classification. SIFT and color features were extracted and a global model was generated using Gaussian Mixture Model (GMM). Bag of Words (BoW) and Fisher Vector (FV) were used to obtain the statistics of feature distributions to train a Support Vector Machine (SVM) classifier.

\begin{figure}[tbp!]
\centering
\includegraphics[width=0.80\linewidth]{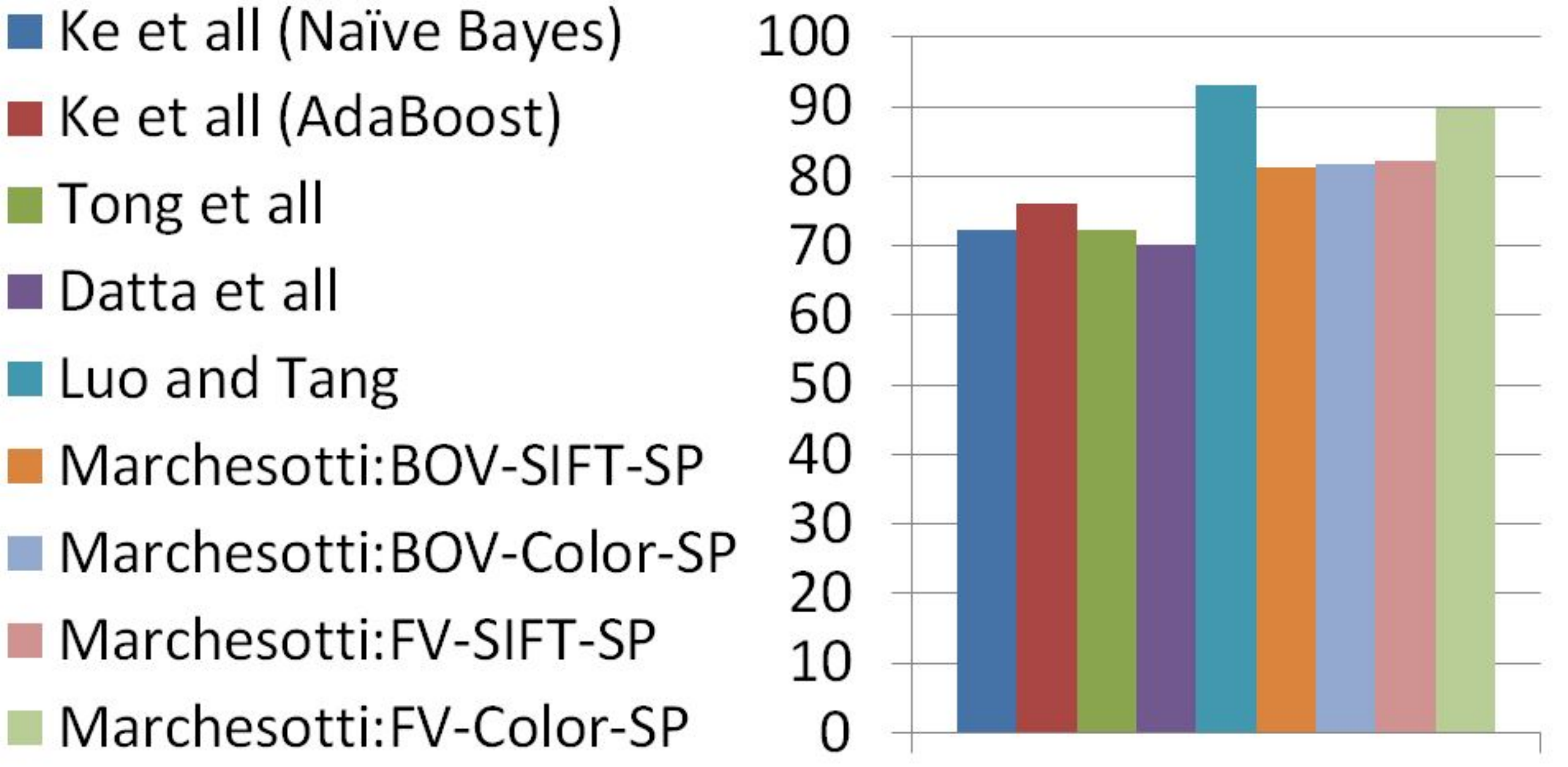}
\vspace{-0.15in}
\caption{Classification accuracy of state of the art methods}
\vspace{-0.2in}
\label{fig:Literature}
\end{figure}

We summarize the classification accuracy of state of the art methods in Table~\ref{fig:Literature}. Hand-crafted global features can reach to a classification accuracy up to $76\%$ with Ke et al. \cite{Ke2006}. Local features proposed by Luo and Tang \cite{Luo2008} leads to an accuracy up to $93\%$. Generic features used by Marchesotti et al. \cite{Marchesotti2011} result in an  accuracy between $81.4\%$ and $89.9\%$. In the rest of the simulations, we use L1 soft-margin SVM classifier. We experimented other classifiers and observed that classification accuracy did not change significantly when we had a large set of training and test images.

\vspace{-0.10in}

\subsection{Geometric Descriptors }
\label{sec:genericgeometric}
GIST is a holistic representation of an image to model the shape of the scene \cite{Oliva2001}. Scene representation is classified with respect to naturalness, openness, roughness, expansion and ruggerdness. GIST uses the local and global energy spectrum to quantify the introduced metrics in the scene representation. We use three different configurations of the GIST features by varying the number of orientations per scale and number of blocks. GIST(16) and GIST(32) correspond to the configuration where we use an unlocalized energy spectra by setting the number of block to \texttt{1}. GIST(512) corresponds to the case where the number of blocks is set to \texttt{4}. Orientations per scales are all set to \texttt{8} for GIST(32) and GIST(512) where orientation scales are set to \texttt{4} for GIST(16). SIFT divides the image into \texttt{4x4} grids of cells and calculates histogram of image gradient directions as explained in \cite{Lowe2004}. GMM approximates the distributions by weighted sum of Gaussian models. In addition to being used as a feature, it is also used to generate the visual dictionary in section \ref{sec:genericvisual}. Maximally stable extremal regions (MSER) thresholds the image in the intensity channel. Threshold is swept from black to white to detect the connected areas that are unchanged over a large set of thresholds \cite{MSER2002}. Difference of Gaussians (DOG) convolves the original image with Gaussian kernels and subtracts the blurred images from each other. Difference of blurred images contains band-pass details that are used as image descriptors. We also detect corners and blobs to represent images using Hessian, Harris and Laplace operators. Implementation of most of the geometric descriptors are provided with VLFeat package which is an open and portable library of computer vision algorithms \cite{Vlfeat2008}.

\begin{figure}[tbp!]
\centering
\includegraphics[width=0.80\linewidth]{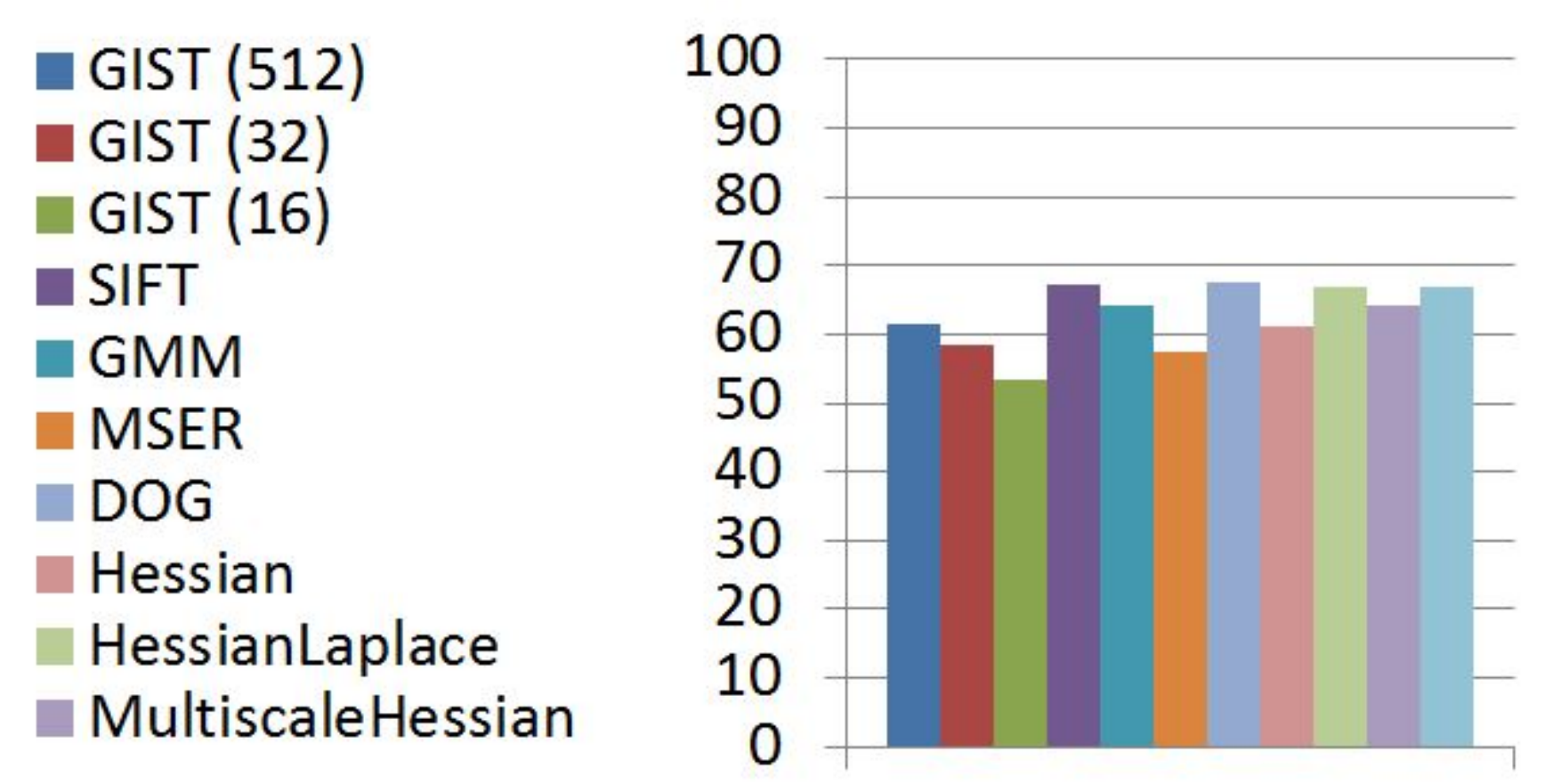}
\vspace{-0.1in}
\caption{Classification accuracy of Geometric Descriptors}
\vspace{-0.2in}
\label{fig:Geometric}

\end{figure}

Classification results for geometric descriptors are given in Figure \ref{fig:Geometric}. In here, we calculate the geometric descriptors by computing the average of each feature dimension over the whole image. We obtain the most accurate classification with $67.7\%$ in DOG, $67.2\%$ in SIFT and $67.1\%$ in HessianLaplace.
Generic geometric descriptors inherently contain information related to the spatial complexity and composition. Some of the geometric descriptors perform better than the others but none of them is sufficient for aesthetics classification since they do not focus on the basic dimensions of the aesthetics.

\vspace{-0.10in}

\subsection{Color Descriptors}
\label{sec:genericcolor}

Color descriptors are designed according to four main constraints: photometric robustness, geometric robustness, photometric stability and generality \cite{Khan2013}. In order to satisfy these constraints, color descriptors should be invariant to shadow, shading, light source configuration, view point, orientation and image quality. However, it is not possible to design descriptors that can satisfy all the constrains. Since image aesthetics is also influenced by these factors, we can use color descriptors as aesthetics metrics.

Color naming is introduced in \cite{ColorNaming2009} as an image descriptor, which calculates the color distributions similar to the bag of words approach. Relative locations of pixels do not effect the distribution since this approach only focuses on the color distribution of the pixels in the region of interest. It is originally used to assign linguistic color labels to image pixels and the main objective is to predict the color category that humans would perceive given a color measurement. In practice, color naming descriptor is a 11-D vector where each dimension corresponds to the distribution of main colors that can be sorted as follows: black, blue, brown, grey, green, orange, pink, purple, red, white and yellow. We also use the color naming method described in \cite{ParametricColorNaming2008}. Authors use a fuzzy $k$-means algorithm to obtain color labels from Munsell book of color. Color descriptor consists of member functions which map color elements to $[0,1]$ interval and the abbreviation JOSA is used to represent this descriptor in the results .

In addition to the distribution of colors, we can also consider the relative locations of the color pixels. We use discriminative color descriptor introduced in \cite{Khan2013} to cluster the color pixels in compact representations. Color pixels are clustered by maximizing the discriminative power using an information theocratical approach. We also use the color descriptors in the opponent color space as described in \cite{Weijer06}. Color descriptors are designed in a similar way to SIFT by calculating the histogram of gradients in hue. The hue color descriptor performs poorly when the saturation is low. In case of low saturation, we can use opponent color angle \cite{Weijer06}.

We use the default version of the color descriptors defined in \cite{ColorCodes}. Color descriptor matrix is composed of three rows where each row corresponds to the metric calculated over three constant regions. In addition, we take the average of the metric over three regions and use it as an additional descriptor which is shown with a suffix $(1x11)$. In case of discriminative color, we also calculate the color descriptors for a color dictionary size of $25$ and $50$.

\vspace{-0.1in}

\begin{figure}[htbp!]
\centering
\includegraphics[width=0.80\linewidth]{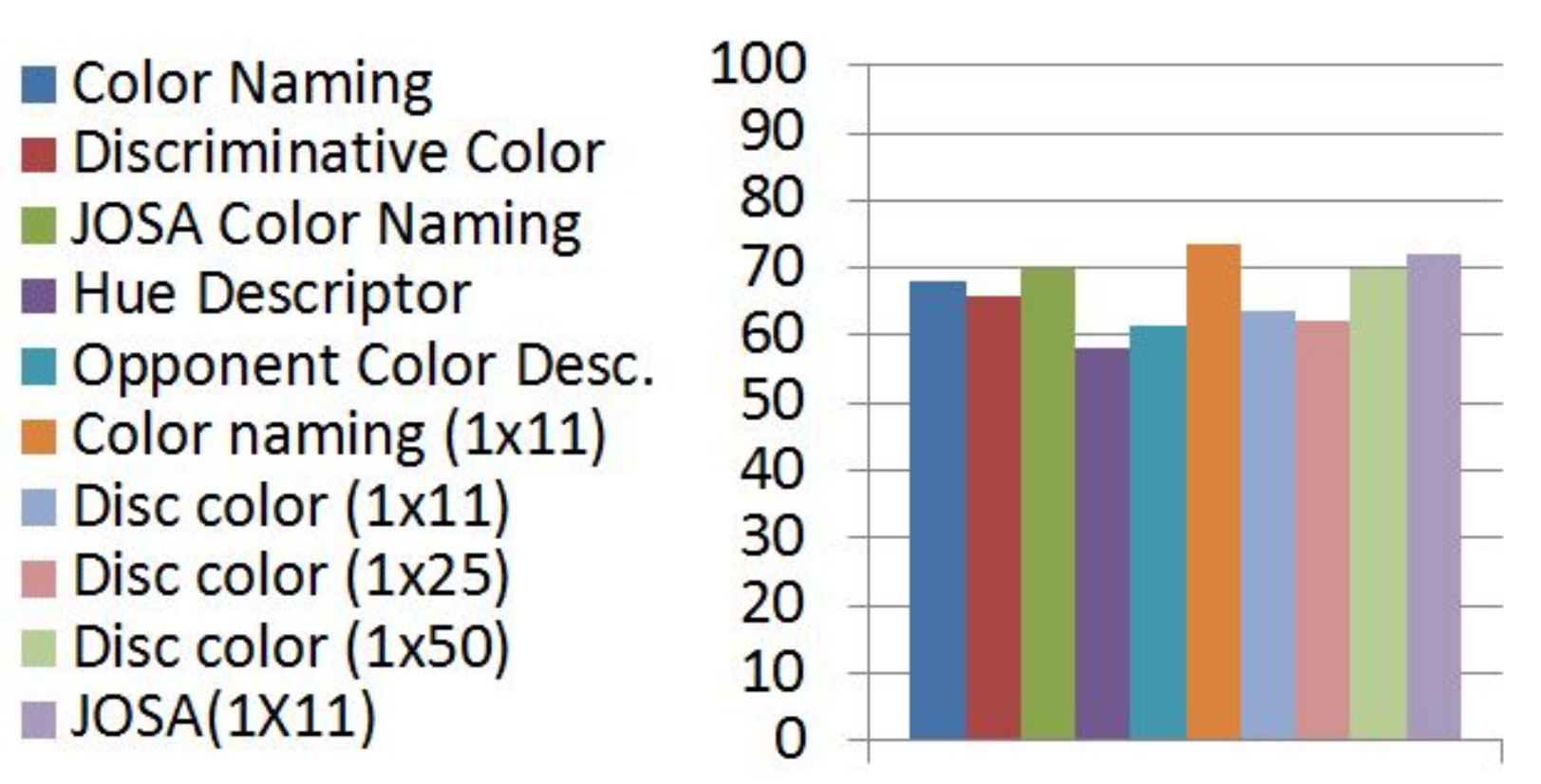}
\vspace{-0.2in}
\caption{Classification accuracy of Color Descriptors}
\label{fig:Color}

\end{figure}

Color naming (1x11) results in the highest classification accuracy with $73.6\%$ followed by JOSA(1x11) with $72.1\%$. When we compare with the generic descriptors, color descriptors perform better. In the literature, variations of color distribution and harmony are commonly used as hand-crafted features to model aesthetics as it is mentioned in section \ref{fig:Literature}.

\vspace{-0.10in}

\subsection{Hybrid Descriptors}
\label{sec:generichybrid}
\begin{figure}[htbp!]
\centering
\includegraphics[width=0.80\linewidth]{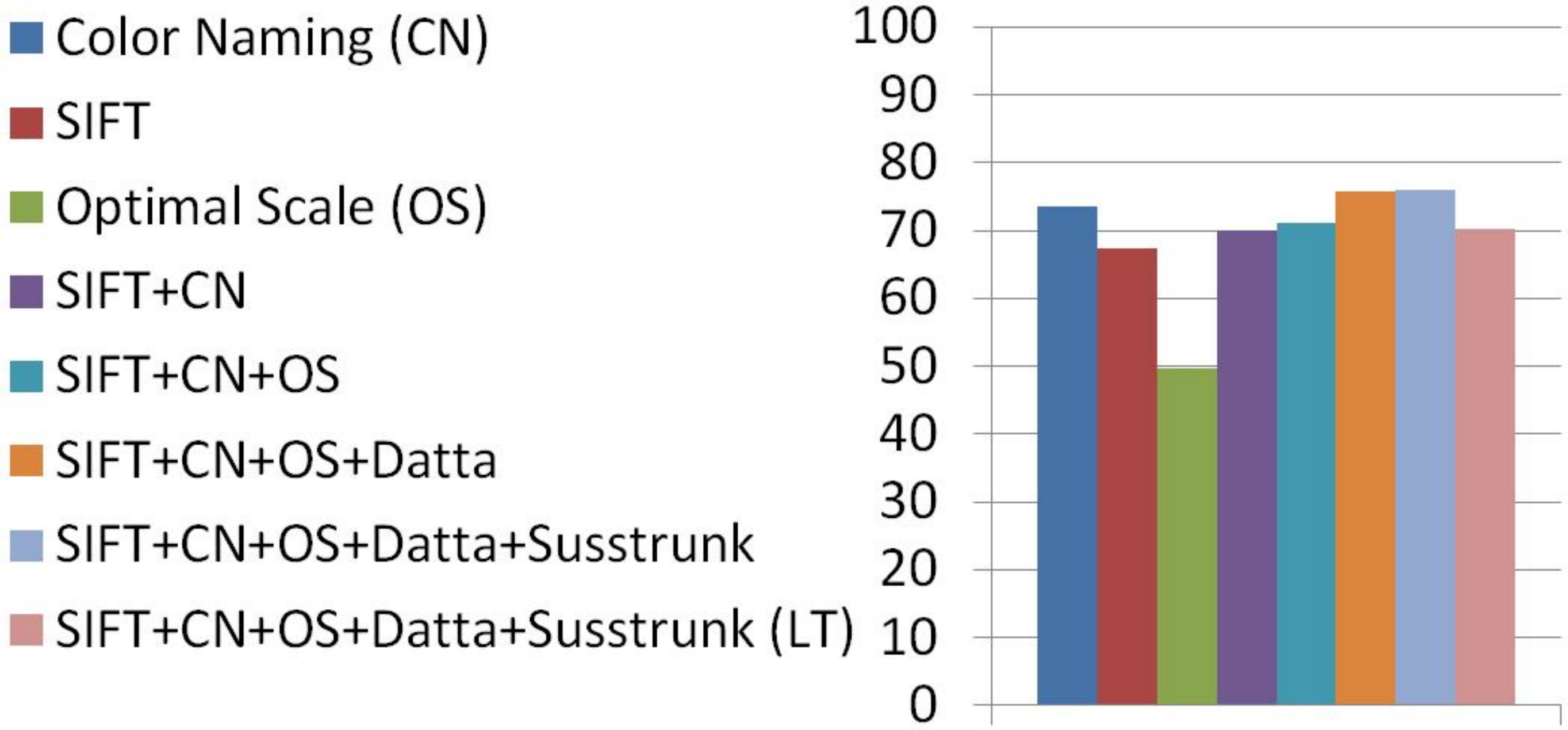}
\vspace{-0.1in}
\caption{Classification accuracy of Hybrid Descriptors}
\label{fig:Hybrid}
\end{figure}

Color, geometric and aesthetics descriptors analyze the images in different aspects and they can be complementary to each other for classification. However, they can also contradict with each other in terms of classification decisions. We combine some of the descriptors defined in the previous parts to obtain a hybrid descriptor. Classification accuracies of the hybrid descriptors are shown in Figure \ref{fig:Hybrid}.

\begin{figure*}[t]
\vspace{-0.1in}
\centering
\includegraphics[width=0.80\linewidth]{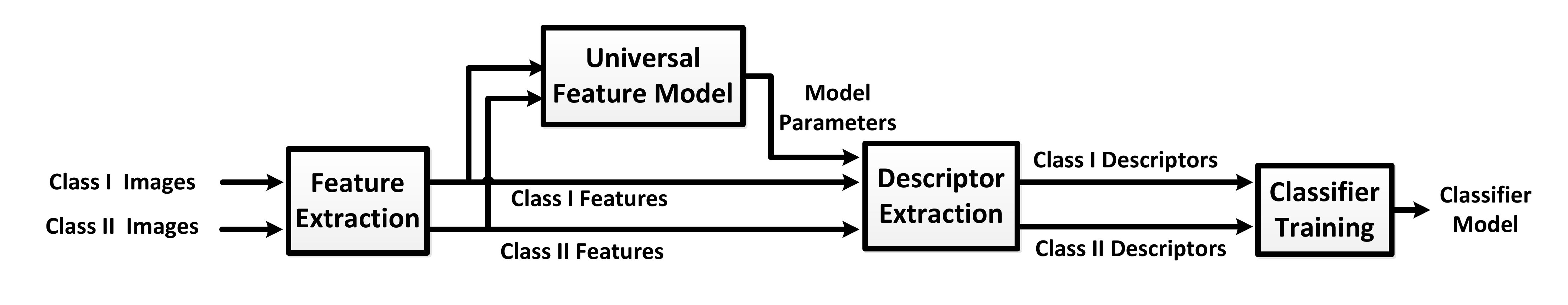}
\vspace{-0.3in}
\caption{Training pipeline}
\label{fig:Training}
\vspace{-0.2in}
\end{figure*}

We use Color naming (1x11) as the color descriptor which is abbreviated as CN and SIFT is used as the geometric descriptor. When CN and SIFT are combined together, classification accuracy gets lower than the accuracy of CN but higher than the accuracy of SIFT with $70.1\%$. Optimal scale (OS) is a feature introduced to quantify the amount of low frequency content in the image. Classification ratio increases by $1\%$ when OS is added to the descriptor. We add exposure of light, hue, rule of thirds, wavelet-based texture and size and aspect ratio features introduced by Datta et al. \cite{Datta2006} to our descriptor. Classification accuracy increases up to $75.7\%$. Finally, we add   brightness, saturation, average color and colorfulness features proposed by Susstrunk et al. \cite{Hasler2003} and \cite{Yildirim2013}. Our hybrid descriptor has a classification accuracy of $75.9\%$. Instead of training with the whole set, we train with the first $100$ good and bad images to experiment the effect of training set size. Classification accuracy drops from $75.9\%$ to $70.2\%$.

\vspace{-0.18in}

\subsection{Visual Dictionary Approach}
\vspace{-0.05in}

\label{sec:genericvisual}

In the previous sections, we extract features and combine them in a vector to form a descriptor. However, we do not use the distribution of the features. In contrast, in this section, we generate a visual dictionary to consider the distributional statistics of the features. The training pipeline including visual dictionary generation is shown in Figure \ref{fig:Training}.

We extract features from two different classes of images and these features are fed to a universal feature model to obtain a visual dictionary. Training features and universal model parameters are fed to a descriptor extraction module to obtain class I and class II descriptors. We train the classifier with the labeled descriptors to obtain a classifier model. We use SIFT, DOG, Color naming (1x11) and JOSA because they have the highest classification accuracy based on our experiments from previous sections. In addition to the default SIFT, we perform singular value decomposition to keep the first $N$ eigenvalues and remove the rest to reconstruct the feature vector. We experimented with different $N$ values and $30$ produced the highest classification accuracy. Gaussian Mixture model (GMM) is used as the universal feature model and Fisher Vector is preferred as the descriptor. In our simulations, we vary the number of Gaussians in the GMM and the training set size. For each feature, we perform simulations at least for $14$ different configurations and up to $22$. Configurations that lead to highest classification accuracy are given in Table \ref{tab:1}.

\begin{table}[htb]
\vspace{-0.10in}
\centering
\caption{Classification accuracy for the dictionary approach  \label{tab:1}}

 \begin{tabular}{|c|c|c|c|} \hline
\textbf{Features}&  \textbf{Number of} & \textbf{Training} & \textbf{Classification } \\
\textbf{Types}&  \textbf{Gaussians} & \textbf{Set Size} & \textbf{Accuracy (\%)} \\\hline
SIFT(SVD) &  \texttt{200} & \texttt{100} & \texttt{75.5}\\\hline
SIFT &  \texttt{200} & \texttt{100} & \texttt{74.0}\\\hline
CN&  \texttt{5} & \texttt{6000} & \texttt{72.6}\\\hline
DoG&  \texttt{2} & \texttt{6000} & \texttt{69.7}\\\hline
JOSA&  \texttt{5} & \texttt{6000} &  \texttt{67.6}\\\hline
\end{tabular}
\vspace{-0.20in}
\end{table}

\vspace{-0.15in}
\subsection{Descriptor Comparison}
\vspace{-0.10in}

\label{sec:comparison}

We observe that state of the art descriptors using local information leads to classification ratios up to $90\%$. Luo and Tong \cite{Luo2008} used hand-crafted features and Marchesotti et al. \cite{Marchesotti2011} used generic features to obtain high classification accuracy. As it was claimed in \cite{Marchesotti2011}, generic features can perform as good as hand-crafted features. However, examining local features in addition to global features have a more significant effect on the classification results than feature selection. The maximum classification ratio we obtain using geometric descriptors is $67.7\%$ and it is $73.6\%$ using color descriptors. Hybrid descriptors lead to $75.9\%$ and the dictionary approach leads to $75.5\%$ at most. Basic generic features such as color and geometric can perform as well as the state of the art global computational approaches. However, they do not perform as well as the ones that take spatial characteristics into account by using subject region extraction or spatial pyramid. The main disadvantage of the regional methods comes from the time and memory complexity. Subject region extraction used in \cite{Luo2008} is an exhaustive approach that requires significant amount of computational time and the spatial pyramid originally introduced in \cite{SpatialPyramid2006} requires significant amount of memory (Approximately 250GB of memory is required to store the extracted features of $1$ dataset out of $4$ in the CUHK dataset ).

\vspace{-0.1in}

\section{Conclusion }
\label{sec:conclusion}
\vspace{-0.1in}

In this paper, we compare generic, hand-crafted and aesthetics descriptors to examine the classification performance of image aesthetics. We have shown that basic generic descriptors can perform as well as the state of the art hand-crafted global descriptors. However, both generic and hand-crafted features are limited in terms of classification when we do not consider the spatial distribution of features. Spatial pyramid and subject region extraction are the main factors that lead to high classification accuracies in the literature. But they require significant amount of computational time and memory. In our future work, we will examine the correlation between the extracted features and overall image aesthetics. Instead of directly feeding the features to classifiers, we will focus on the individual relationships between the features and aesthetics to model a no-reference image aesthetics metric based on deep learning. We plan to use AVA dataset introduced in \cite{Murray2012} to evaluate the aesthetics metrics because of the image variety, aesthetics scores and rich annotations.

{

\end{document}